\documentstyle[12pt]{article}
\newcommand{\be}{\begin{equation}}
\newcommand{\ee}{\end{equation}}
\newcommand{\bea}{\begin{eqnarray}}
\newcommand{\ena}{\end{eqnarray}}
\def\ep{\epsilon}
\begin{document}
\begin{center}
\vspace{1cm}
{\large \bf The Retarded-Advanced basis in the Real Time
formalism\footnote{presented at the 3rd Workshop on Thermal Field
Theory, Banff, August 1993}}
\bigskip

 { F.GUERIN} \\ {\em Institut Non Lineaire, UMR 129 du CNRS, 1361
Route des Lucioles \\ 06560 Valbonne, France}

\end{center}

\section{Introduction}

In the real time formulation of thermal field theory, there is a
doubling of the number of degrees of freedom, a "ghost" field (type
2 field) is associated to each physical field (type 1 field).
Aurenche and Becherrawy \cite{Aur} have reformulated the real time
perturbation theory in terms of retarded (R) and advanced (A)
propagators. The usual propagator in the (1,2) basis is expressed
in terms of retarded, advanced ones
\be
\Delta_{R,A}(p)={i\over{(p_0+i\ep_p)^2-{\bf p}^2-m^2}}  \  \  \

{\rm with} \ \ \  \ep_p>0 \ \ {\rm for\  R},\  \ \ep_p<0 \ \

{\rm for \ A} \ee
\begin{eqnarray}
U {\left( \begin{array}{cc} {\Delta_{11}} & {\Delta_{12}} \\

{\Delta_{21}} & {\Delta_{22}} \end{array}\right) } V = {\left(
\begin{array}{cc} \Delta_R & 0 \\ 0 & \Delta_A \end{array} \right)}
\end{eqnarray}
and the transformation matrices $U$ and $V$ are moved
from the propagator to the vertices. As a result, one is led to new
types of vertices in the R,A basis \cite{Aur,Wee}. For example, in a
$\phi^3$ theory there are two bare vertices
\be
\gamma_{RRA}= \ ig\ \ \ \ \ {\rm and} \ \ \ \ \
\gamma_{AAR}(p,q,r)= \ -ig \ \ N(p,q)
\ee
\be
N(p,q) = n(p_0)+n(q_0)+1 = n(p_0) n(q_0)  /  n(p_0+q_0)
\ee
where the momenta are incoming the vertex and $n(p_0)$ is the
Bose-Einstein factor. In other words, the propagators are the $T=0$
 ones, and the $T$ dependence is in the statistical weight
$N(p,q)$ associated to one bare vertex (an alternative choice for
the $U,V$ matrices reverses the role of R and A). General
relations exist between the amputated Green functions
\cite{Wee,Aur}; with all momenta incoming, they are
\be
\Gamma (p_{1 A}, \ldots , p_{n A}, q_R) = N(p_1,\ldots, p_n) \ \
\Gamma^{\ *} (p_{1 R}, \ldots , p_{n R}, q_A)
\ee \be
\Gamma (p_{1 A}, \ldots , p_{n A}, q_{1 R}, \ldots ,q_{m R})
 = {N(p_1,\ldots,p_n)\over{N(q_1,\ldots,q_m)}} \

\Gamma^{\ *} (p_{1 R}, \ldots , p_{n R}, q_{1 A}, \ldots , q_{m A})

\ee \be
N(p_1, \ldots, p_n) = \prod_{i=1}^n n(p_i^0) \ / \ n(\sum_{i=1}^n
p_i^0) = \prod_{i=1}^n (1+n(p_i^0)) - \prod_{i=1}^n n(p_i^0)
\ee
and the dressed propagators  obey
\be
\Delta_R^d(p) = \Delta_A^d(-p)
\ee
The generalisation to the case of mixed fermions and bosons is
simple, $n(p_i^0)$ is replaced by $\eta_i n(p_i^0-\mu_i)$, with

$\eta_i=+1$ for bosons, $\eta_i=-1$ for fermions, $\mu_i$ is the
chemical potential, and $ S_{R,A}(p) = (\gamma . p + m)
\Delta_{R,A} (p) $.

\bigskip

For multiloop diagrams, the use of this basis looks, at first,
not to improve on the (1,2) basis, since one has to sum over all
possible cases for the $R,A$ components of the internal lines.
However the result of this summation turns out to be simple and
general, so that the tedious summation is avoided. Indeed, the
$R,A$ Green fuctions are linearly related \cite{Eva} to boundary
values of analytical continuations of the imaginary time (IT) Green
functions. In  the IT formulation, the Feynman rules are in
terms of imaginary energies and one single component field.

The simplest example is
\be
\Gamma(p_{1 R}, p_{2 R}, \ldots, p_{n R}, q_A) =
\Gamma^{IT}(p_1^0+i\ep_1, p_2^0+i\ep_2, \ldots, p_n^0+i\ep_n,
q^0+i\ep_q)
\ee
with $\ep_i>0 , \ep_q<0 ,\ \  \sum_{i=1}^n\ep_i + \ep_q = 0 $. More
generally, for any N-point IT Green function, the relevant analytical
continuations of the external energies $p_i^0$ are such that

$\sum_{i=1}^N p_i^0 =0$ both for the real and imaginary parts,
i.e. one has $\sum_{i=1}^N\ep_i=0$
\section{Perturbative properties of R,A Green functions}
The R,A Green functions share the good properties of the IT
amplitudes, and possess additional welcome features \cite{Fra}. For
any  multiloop diagram

a) the amplitude may be written as a sum over all the possible tree
diagrams obtained by cutting internal lines; the weight associated
to a cut line of momentum $k$ is
\ \ \ \ $(n(k_o)+1/2)(\Delta_R(k)-\Delta_A(k))$ , with \be

\Delta_R(k)-\Delta_A(k) = 2\pi \ep(k_0) \delta (k_0^2-{\bf k}^2-m^2)
\ee

b) the Cutkosky rules are very similar to the $T=0$ case : the
intermediate states are defined in the same way and weighted by
$N(p_1, \ldots, p_n)$ ; factorisation properties hold.

c) Properties (a),(b) are valid for the case of diagrams made of
dressed propagators and bare vertices, the weight for a "cut" line
now involves $(\Delta_R^d(k)-\Delta_A^d(k))$ i.e. the spectral
function. If that function is dominated by a quasi particle, one
has a diagrammatic interpretation of the amplitude.

d) The constraints from the external energies' epsilons are simply
accommodated if one defines a new quantity, the $\ep$-$flow$. Indeed
one may keep track of the $R,A$ components in a graphical way,
similar to the components of a charged scalar field, i.e. one
defines an arrow associated to the $\ep$-flow.\\ For all momenta
incoming the Green functions,  $R$ lines have incoming
$\ep$-flows, $A$ lines have outgoing ones ; the flow is conserved
at each vertex and, as a result, it obeys Kirchoff laws. This
property allows simple graphical rules when a diagram is written
as a sum of trees. For example, for the simple case of Eq.(9),
the $\ep$-flow along each tree is determined, as there is one sink
and several sources.

A simple example is the one-loop amplitude for the vertex
$\Gamma_{RRA}$. The summation over all possible $R,A$ components
 around the loop produces four  loop configurations, with
Feynman rules given by Eq.(3). That sum may be rewritten as the
sum over the three possible trees and the $\ep$-flow is fixed
along the trees' propagators. The generalisation to the case of
dressed propagators and bare vertices is straightforward, thanks to
the $\ep$-flow.

\bigskip
At this point it is interesting to look at the number of {\it
degrees of freedom}. In the complex $p_0$ plane, the $R$ or $A$
propagator possesses two poles on the same side of the real axis,
while the Feynman propagator has one pole on each side. At $T=0$,
in the S-matrix elements, the incoming fields are associated to
positive on-shell energies and the outgoing fields to negative
on-shell energies. Those are the physical fields in the $R,A$
basis (as well as in the Feynman one), and the ghost fields are
associated to the opposite prescription. In a loop summation,
physical and ghost fields cooperate to produce the factor
$\Delta_R(k)-\Delta_A(k)$ in property (a).

\section{The linear relations between $R,A$ and IT Green
functions}

As a foreword, a general property is recalled. To each possible
analytic continuation of an IT N-point Green function (subjected
to $\sum_ip_i^0=0$) is associated a "generalized" retarded
function. Those have been defined in 1961, mostly by Araki and
Ruelle \cite{Ara}, in connection with the proof of the crossing
properties of S-matrix elements. Those authors constructed a full
set of retarded products of operators, whose matrix elements are
analytic in one domain such that all energy sums (the imaginary
parts) have a fixed sign. The construction for bosons follows, for a
given domain. Let a time ordering $t_a, t_b, t_c, \ldots$ be:\\  i)
the operators are ordered according to time order, ii) the momenta
conjuguate to the time intervals are defined \\
$p_at_a+p_bt_b+p_ct_c+\ldots =
p_a(t_a-t_b)+(p_a+p_b)(t_b-t_c)+(p_a+p_b+p_c)(t_c-t_d)+\ldots$ \\
iii) to each time interval $(t_i-t_{i+1})$, one associates a factor
$\theta(t_i-t_{i+1})$ if the conjuguate momenta $P_i$ is such that
Im $P_i>0$, and a factor $-\theta(t_{i+1}-t_i)$ if Im $P_i<0$ . One
then sums  over all time permutations.

 \bigskip

The linear relation between $R,A$ and IT Green functions is generally
more complicated than Eq.(9). The case of the 4-point function is
chosen as an example. The amplitude $\Gamma(p_A,q_A,r_R,s_R)$ is
related to the analytical continuations of the IT amplitude \\

$\Gamma^{IT}(p_0+\ep_p, q_0+\ep_q, r_0+\ep_r, s_0+\ep_s)$, \ \  such
that \ \ $\ep_p,\ep_q>0$ and $\ep_r,\ep_s<0$ with
$\ep_p+\ep_q+\ep_r+\ep_s=0$.
 \\ There are two subenergies $p_0+r_0$ and $p_0+s_0$ for which the
analytical continuation is unconstrained, as $\ep_p+\ep_r$ and
$\ep_p+\ep_s$ may be of either sign. As a consequence, there are
four analytic continuations linked by one relation. Defining
\be
\Gamma_{++}=\Gamma^{IT}(\ep_p+\ep_r >0, \ep_p+\ep_s >0 ) \ \ \ ,

\ \ \  \Gamma_{++} + \Gamma_{--} = \Gamma_{+-} + \Gamma_{-+} \ \
\ \ \ee is the relation. It is an identity of the
generalized retarded functions \cite{Ara}. Alternatively this
relation is immediate in perturbation theory, as the amplitude is
the sum of a function of $p+r$ and of a function of $p+s$; indeed
each tree may depend either on one or on the other.

The amplitude $\Gamma(p_R, q_R, r_A, s_A)$ is a specific linear
combination of those four amplitudes, determined by the Feynman
rules (Eq.(3)). A systematic feature of the R,A basis is that one has
to sum over all the possible cases for the $\ep$-flow (with an
appropriate weight), once the diagram's amplitude is written as a sum
of trees. In the present example, the trees that depend on $p+r$
 carry a weight $N(r, q+s)$ for the case $\ep_p+\ep_r>0$, and a
weight $N(s, p+r)$ for the case $\ep_p+\ep_r<0$. One resulting
form that generalizes to an N-point function is
\be
\Gamma(p_R, q_R, r_A, s_A) = N(r,s) \Gamma_{++} + N(s,p+r)
(\Gamma_{-+}-\Gamma_{++}) + N(r, p+s) (\Gamma_{+-}-\Gamma_{++})
\ee
i.e. the amplitude is written as one  IT analytic continuation

 plus two discontinuities in only one channel, $p+r$ or $p+s$.
Those discontinuities obey Cutkosky  rules similar to the
self energy case; alternatively they have general forms in terms of
commutators of operators products \cite{Ara}.

Similarly , the 5-point amplitude $\Gamma(p_R, q_R, r_A, s_A,
t_A)$ is the sum of one IT analytic continuation, weighted by
$N(r,s,t)$, plus six discontinuities in one of the subenergies
$p+r, p+s, p+t, q+r, q+s, q+t$. More generally, the useful set of
subenergies are the multiperipheral momentum transfer variables, if
$p+q$ is the incoming channel.

\section{Evaluation of discontinuities in the R,A basis}

The
definition of the possible intermediate states associated to a
diagram is the same as for $T=0$, i.e. the cutting plane splits the
diagram into two, and only two, pieces (in contrast to the (1,2)
basis or the old time-ordered perturbation theory).

 An n-particle intermediate state is weighted by $N(q_1,\ldots,
q_n)$, as given by Eq.(7).

For example for a diagram contributing to the self energy,

\be
\Sigma(p_0+i0) - \Sigma(p_0-i0) = \sum_n  \Gamma^{(1)}(p_A, q_{1 R},
\ldots, q_{n R}) \ \Gamma^{(2)}( (-q_1)_R, \ldots, (-q_n)_R,
(-p)_A ) \ee
\be
\sum_n = \sum_n \int \prod_{i=1}^n (d_4q_i
[\Delta_R(q_i)-\Delta_A(q_i)] )\ \  \ N(q_1, \ldots, q_n)
\ee
and the sum is over all possible intermediate states. An
equivalent form exists where the roles of R and A are
interchanged in Eq.(13). Thanks to the $\ep$-flow, the property
generalizes to diagrams made of bare vertices and dressed
propagators. \\ More precisely, if a multiloop diagram is written as
a sum of trees, the left member of Eq.(13) prescribes to evaluate the
difference between the two orientations of the $\ep$-flow along each
tree; the result is a $\Delta_R-\Delta_A$ factor associated to
one of the tree's internal lines, i.e. the tree is split into two
parts, in all possible ways. Collecting all pieces, the
amplitudes on each side of the cutting plane are written as a sum
of trees in Eq.(13).
\medskip

In a similar way, the discontinuity across one multiperipheral
momentum transfer variable involves the difference between the two
possible orientations of the $\ep$-flow along that momentum, all
other compatible transfer momenta being fixed. For the example of the
4-point function considered in Eq.(12), $\Gamma_{++} - \Gamma_{-+}$
is the discontinuity across the $p+r$ channel, and one has
\be
\Gamma_{++} - \Gamma_{-+} = \sum_n \Gamma^{(1) 2+n}(p_A, r_R,
q_{1 R}, \ldots, q_{n R}) \Gamma^{(2) 2+n}( (-q_1)_R,\ldots,
(-q_n)_R, q_A, s_R)
\ee
with $\sum_n$ given by Eq.(14) and the sum is over all possible
ways of cutting the diagram into two pieces in the $p+r$
variable. (An equivalent form exists with all $q_{i A}$).
\medskip

The 3-point vertex is another example; it is interesting to compare
the different functions, in connection to the definition of an
effective coupling constant. The quantity

\\ $\Gamma(p_R, q_R, r_A) + \Gamma(p_A, q_R, r_A)  N^{-1}(p,r) =
D(q_R, r_A)$ \\

compares the $\ep$-flow when $p$ is a source or $p$ is a sink. In
this difference, a diagram is split into two pieces where the
external leg $p$ is on one side of the cutting plane, and $q, r$ on
the other side. One now notices that  \\ $D(q_R, r_A) - D(q_A, r_R)
= 2 \ {\rm Im} ( \Gamma(p_R, q_R, r_A) - \Gamma(p_R, q_A, r_R) ) = i
X $ \\
In that difference, a second cutting plane shows up, with $q$ on one
side and $r$ on the other one, and $X$ is  written as a sum of
double discontinuities. As a consequence,  the difference
between the real parts of the two effective coupling constants may
only show up in processes that are sensitive to those double
discontinuities.

\bigskip
To {\it conclude}, the $R,A$ basis provides the precise link
between the imaginary time Green functions, and the Green function
in the usual (1,2) basis of the real time formalism. Simple
diagrammatic rules exist, where the spectral function appears
naturally in diagrams made of dressed propagators and bare vertices.
\bigskip

{\bf Acknowledgments}
\\
I wish to thank R. Stora for sharing a bit of his knowledge. A
collaboration with P.Aurenche lead to the quasiparticle aspect and
the identification of the degrees of freedom.


\begin{thebibliography}{99}

\bibitem{Aur} P. Aurenche and T. Becherrawy, {\it Nucl. Phys.}
{\bf B379} (1992) 259.


\bibitem{Wee} M.A. van Eijk and Ch.G. van Weert, {\it Phys. Lett.}
{\bf B 278} (1992) 305.

 \bibitem{Eva} T. Evans, {\it Nucl. Phys.} {\bf B374} (1992) 340.


\bibitem{Fra} F. Guerin, INLN preprint 1993/14 (hep-ph/9306216)

\bibitem{Ara} H. Araki, {\it J.M.P.} {\bf 2} (1961) 163 ; D.
Ruelle, {\it Il Nuov. Cim.} {\bf 19} (1961) 356 ; R. Stora, p.28 in
{\it Les Houches 1971}, ed. Gordon and Breach (New York 1973)

\end{thebibliography}
\end{document}